# 4f fine-structure levels as the dominant error in the electronic structures of binary lanthanide oxides

Bolong Huang*

*Department of Physics and Materials Science, City University of Hong Kong, Kowloon, Hong Kong SAR, China*
*The Present Address: Department of Applied Biology and Chemical Technology, The Hong Kong Polytechnic University, Hung Hom, Kowloon, Hong Kong SAR, China*

*Email: bhuang@polyu.edu.hk

## Abstract

The ground-state 4f fine-structure levels in the intrinsic optical transition gaps between the 2p and 5d orbitals of lanthanide sesquioxides ($Ln_2O_3$, Ln=La…Lu) were calculated by a two-way crossover search for the U parameters for DFT+U calculations. The original 4f-shell potential perturbation in the linear response method were reformulated within the constraint volume of the given solids. The band structures were also calculated. This method yields nearly constant optical transition gaps between Ln-5d and O-2p orbitals, with magnitudes of 5.3~5.5 eV. This result verifies that the error in the band structure calculations for $Ln_2O_3$ is dominated by the inaccuracies in the predicted 4f levels in the 2p-5d transition gaps, which strongly and non-linearly depend on the on-site Hubbard U. The relationship between the 4f occupancies and Hubbard U is non-monotonic and is entirely different from that for materials with 3d or 4d orbitals, such as transition metal oxides. This new linear response DFT+U method can provide a simpler understanding of the electronic structure of $Ln_2O_3$ and enables a quick examination of the electronic structures of lanthanide solids prior to hybrid functional or GW calculations.

## Introduction

Solids based on the f-electron-containing rare earth lanthanide (4f) and actinide (5f) elements are widely used in such applications as catalysis, luminescence (up-conversion), oxygen and hydrogen storage and permanent magnets. Their advantageous properties for these applications are closely related to the complicated low energy differences between the f-electron levels in either the ground or excited states. Each of the lanthanide and actinide elements in the periodic table has attracted considerable interest in the past decades.

For instance, materials with fully filled 4f orbitals, such as $Lu_2O_3$, are good candidates for super-persistent red and green luminescent/phosphor applications when doped with $Eu^{3+}$ and $Tb^{3+}$.[1,2] As is well known, the major application of these materials is in lasers and as phosphors, which requires the precise electronic spectra of lanthanide ions ($Ln^{3+}$). Knowledge of these spectra requires a high degree of accuracy for both experimental instruments and theoretical calculation methodologies, as the assignment of the $4f^N$ energy levels are described by the J-multiplets $^{2S+1}L_J$ (recalling the Russell-Saunders scheme), where S, L and J denote the electron spin, orbital, and total (i.e., vector sum of S and L) angular momenta, respectively in units of the Plank constant $\hbar$. For each J value, there are (2J+1) possible microstates.[3] For example, for $Ce^{3+}$, $4f^1$, there are two multiplets, $^2F_{5/2}$ and $^2F_{7/2}$ (where S = 1/2, L = 3, and J = 5/2 or 7/2), and for $Eu^{3+}$, the lowest multiplet is $^7F_J$ (J = 0–6), where S = 3, L = 3 and J can take integral values between 0 and 6, with J = 0 being the ground state.[3] Therefore, we call the $4f^N$ energy levels the 4f fine-structure levels (or 4f fine levels). Based on band theory, the 4f fine levels are assigned variously below, above and within the 2p-5d optical transition gap in lanthanide oxides across the periodic table from La to Lu.

Elucidation of the electronic properties of these materials using theoretical methods is challenging because most of these materials are mixed valence or heavy fermion systems.[4,5] In addition to exhibiting strong correlation, the f-electrons also couple with the s, p, and d (mainly 5d) electrons. This imposes stringent requirements for an accurate description of the electronic states. In this work, to treat the electronic structures of f-electron solids using a simpler model, we focus on the relative energy differences between the 4f energy levels and the occupied valence band maximum (VBM), which is one of the possible sources of error in density functional theory (DFT) electronic structure calculations for 4f- or 5f-based solids. We use lanthanide oxides as an example to demonstrate our new linear response method, which improves the accuracy of the 4f orbital energies using a simple DFT+Hubbard U technique.

The electronic properties of lanthanide sesquioxide ($Ln_2O_3$, Ln=La,…, Lu) have been well studied, with Prokofiev et al demonstrating a periodic-like variation of the optical band gaps.[6] However, local density approximation (LDA) or generalized gradient approximation (GGA) has been found to vastly underestimate the band gaps,[7,8] and the results obtained using the DFT+U method with U for 4f orbitals only deviate greatly from the experimental values.[9-11] The origin of this difference may be the incomplete counteraction of self-interaction induced by the localized hole states, as mentioned in previous work.[12] Advanced methods, such as GW[13,14] or hybrid functionals[15] based on fixed experimental lattice constants, substantially improve band structure calculations and reproduce the variation behavior found by Prokofiev et al, as reported in our previous work and that of Jiang et al.[13,14] Additionally, our previous work found that the use of such methods as hybrid functionals in conjunction with pseudopotential methods can also introduce errors in the DFT-predicted lattice parameters and defect levels.[12,16]

Another issue is whether pseudopotential-based DFT can describe the $4f^N$ spectra of lanthanides. In LDA/GGA, the ultrasoft or normconserving pseudopotential cannot distinguish between $4f^N$ and $4f^{N-1}5d$ well. This shortcoming is a result of the fact that the current self-consistent pseudopotential is established by an orbital eigen energy that is relatively low but similar to that of the 5s-5p-5d-6s shells. Thus, the valence charge density does exhibit considerable overlaps, especially between the 5d-6s orbitals. Although the non-linear core correction helps to screen the core-valence charge density, the 4f orbital is difficult to distinguish energetically within valence charge densities.

It is well known that the band gaps of approximately 5.5-6.0 eV for $Ln_2O_3$ materials are predominantly due to optical transitions from the O 2p to Ln 5d orbitals.[6,17-20] We also found that the lattice parameters predicted using either local density approximation or generalized gradient approximation (LDA/GGA) or DFT+U levels are essentially identical. An example of this similarity can be found for $CeO_2$ in our previous work.[12] Moreover, Lany and Zunger have shown that the localized hole states are actually induced by O or similar anion elements of d-orbital based compounds;[21,22] this was confirmed by Keating et al for f-orbital based materials[23] and by our previous work.[12] The basis of the present work is the finding that the localized hole states from 2p orbitals induced nearly the same repulsive Coulomb potential of approximately 5.5-5.6 eV, and the localized 5d orbitals also exhibit nearly constant Coulomb potentials of approximately 2 eV. This leads us to believe that the 2p-5d gaps of $Ln_2O_3$ are actually constant optical transitions regardless of both the identity of the elements in $Ln_2O_3$ and the lattice parameters of the particular structures. Recalling the previous hybrid functional calculations[15] and the GW work by Jiang et al,[13,14] these DFT+Hubbard U results further support this deduction. In the present work, we attribute the main errors in the electronic structure calculations to the 4f relative energy levels at the top of occupied valence band.

To predict the on-site Coulomb potential energy (Hubbard U), one can simply use a Cococcioni-type linear response method to obtain the localized f orbitals.[24] However, the strong coupling between the 4f and 5d-6s orbitals leads to unintentional overlapping, such that the difference in the electron wavefunction occupancies contains not only the 4f electrons but also overlaps with a portion of the anti-directional perturbed 5d-6s orbitals. This is due to similar orbital eigen energies among 4f, 5d, and 6s orbitals, and the perturbed charge occupation of 4f orbitals in practical linear response processes consequently follows $\Delta n = \Delta n_{4f} + \Delta n_{5d} + \Delta n_{6s}$. As observed in this relationship, because the practical $\Delta n > \Delta n_{4f}$, the $(\alpha_p / \Delta n) < (\alpha_p / \Delta n_{4f})$ or $U_{calculated} < U_{real}$ for a given orbital potential perturbation $\alpha_p$ in conventional linear response method (explaining why the 4f electrons are screened in the solids), giving a large error in the Hubbard U predictions. Thus, the simple estimation of the U parameter through the inverse of the small perturbation by linear response yields unphysical values. This also results in band gap values that are either too small or too large for both partially and fully filled 4f solids. This phenomenon can be seen in the data summarized in Table 6 in the book chapter by Bendavid et al.[25] The values show that the U for the f orbital has been fixed at approximately 5 eV, while the resulting electronic band gaps are in the 2-3 eV range, showing an approximately 40% error compared to the experimental data.[6] Thus, the DFT+U formalism can give nearly self-interaction-free results if the charge occupation overlapping induced potential relaxation and self-energy of the originally given charge occupation on the targeted orbitals can be well offset or quantitatively described. In this case, both the forward (orbital relaxation) and backward (original self-energy) errors are counteracted in the linear response process. We therefore reformulate the procedure of Cococcioni et al[24] for the determination of the U parameter by linear response and explore the intrinsic properties of self-consistent Hubbard U determination calculations.

## Methodology

### (1) Physical model

As described above, the Hubbard U for f-electrons cannot be simply predicted by the one-way Cococcioni linear response method due to the spontaneous overlap of the 5d and 6s orbitals. To address this issue, we use a two-way perturbation within a certain volume V of the solids. The nearby orbital spontaneous charge overlapping is a systematic error that rigorously exists for any given charge occupancy level for a one-way linear response. This is actually a complicated physical effect in which the polarization of the anions by the 4*f* or 5*d* valence electron and dynamical correlation with the valence electrons of the anions both contribute to the shift. This shift has been found to be difficult to calculate *ab initio.* Dorenbos and coworkers have explored the effects using a Hartree-Fock Møller-Plesset method.[26] However, we focus on a simplified method based on DFT for the orbitals in this work.

For partially filled f-orbital lanthanide elements, especially for light lanthanides from La to Eu, the large fraction of empty components of f-orbital projections easily accommodates the fluctuating f electrons. Simply using a U for a linear response disturbs the highly sensitive original charge occupancy due to the self-induced dipole effect of dynamic polarization with neighboring ions.[27] Therefore, the essential mechanism is due to the relaxation-polarization from the residual orbital self-energy arising from the improper choice of U.

Using the approximation of Zunger and Freeman,[21,28] we updated the realistic electron addition energy in the current DFT+U theory scheme according to

$$E(N+1)-E(N)=e_i(N)+\sum\nolimits_i(U)+\Re_i(U) \tag{1}$$

where $e_i(N)$ are the *i*th orbital eigenvalues of the N-electron system, $\sum_i(U)$ is the spurious self-energy of the *i*th orbital related to U, and $\Re_i(U)$ contributes the orbital wavefunction relaxation energy in excited states, which can be simplified as the difference between the relaxed and unrelaxed self-consistent excitation energy solutions in the DFT+U scheme. This contribution is treated as the relaxation-polarization of self-energy at the *i*th orbital mentioned above. $\sum_i(U)+\Re_i(U)$ has a nonlinear relationship with the U correction, as empirical tuning of U for correcting the orbital eigenvalues $e_i(N)$ led to a non-zero contribution of the $\sum_i(U)+\Re_i(U)$ such that the Janak theorem[29] has a inequality. Solving this relation requires considerable mathematical effort; therefore, we choose a simplified physical model.

We accordingly treat the self-energy and orbital wavefunction relaxation as functionals of the charge density of the system such that $E[\rho]=E[\rho_0]+E_{SE}[\rho]+E_R[\rho]$, where $E_{SE}$ and $E_R$ are the self-energy and orbital wavefunction relaxation, respectively; $\rho$ is the constructed charge density based on Fermi-Dirac distribution; and $E[\rho_0]$ is the total energy of the system in the real world with the true charge density.

In another publication, we provided a simplified model of the self-energy and orbital wavefunction relaxation in the orbitals. We proposed a simple physical idea that, *the perturbations of self-energy and wavefunction relaxation with related to the occupation number in the orbital should follow the behavior of Hubbard-type interaction, regardless the absolute values of both.* We approximate this model within the framework of local density (LD) formalism and model the self-energy projected on the orbital by components as follows:

$$E_{SE}[\rho]=\sum_i n_i\sigma_i \text{ and } E_R[\rho]=\sum_j n_j r_j \tag{2}$$

The wavefunction has to be orthonormal by a linear combination of atomic orbital (LCAO). This pre-condition is easily achieved by the packages VASP, CASTEP, or PWScf. This model is advantageous because the on-site orbital self-energy projection on the other orbital is zero, i.e., the self-energies on the orbitals do not affect one another. To maintain the Janak theorem,[29]

$$\frac{\partial E[\rho]}{\partial n_i}=\varepsilon_i=\varepsilon_{i,real}+\sigma_i+r_i \tag{3}$$

Because we can choose a simplified model with only one orbital component in an atomic site, the total energy functional to the orbital charge occupation of the $I$th atomic site is

$$E[n_I] = \min[E[\rho] - \sum_I (n_I \sigma_I + n_I r_I)] \tag{4}$$

To solve the Eq. (4) by DFT using the one-electron Kohn-Sham equation, we have to choose a perturbation method. According to the approximation in Eq. (4) (i.e., no crystal-field splitting effect), the on-site orbital self-energy part can be expanded as a perturbation, $\sigma_I = \sigma_I(0) + o(\sigma_I) = \sigma_I(0) + \alpha_I$. The term $\sigma_I(0)$ is the $I$th atomic site original self-energy potential induced by the localized orbital, and $o(\sigma_I)$ or $\alpha_I$ is the first-order perturbation of the eigenvalue of the on-site self-energy $\sigma_I$. We model the orbital wavefunction relaxation in terms of the change of occupation from $n_I$ to $(n_I + \delta n_I)$ under the LDA+ $U_I$ local potential environment. By Maclaurin series expansion (a simplified form of Taylor expansion), we obtain

$$n_i(\alpha_i) = n_i(0) + \frac{\partial n_i}{\partial \alpha_i} \cdot \alpha_i + o\left(\frac{\partial n_i}{\partial \alpha_i} \cdot \alpha_i\right) \cong n_i(0) + \left(-\frac{\alpha_i}{U}\right) \tag{5}$$

The above term $\delta n_i$ is therefore described by $\delta n_i = n_i(\alpha_i) - n_i(0) = (-\frac{\alpha_i}{U})$, acting as a Lagrange multiplier. The one-electron effective self-energy potential and relaxation potential are approximated as quadratic functions of the self-energy and orbital relaxation according to Eq. (4):

$$\begin{aligned} U_I + (U_{SE} - U_R)_I &= -\left( \frac{\partial^2 E[n_I]}{\partial n_I^2}\Big|_{KS-SCF} - \frac{\partial^2 E[n_I]}{\partial n_I^2}\Big|_{initial} \right) \\ &= \left(\frac{\partial \sigma_I}{\partial n_{I,\sigma}} + \frac{\partial r_I}{\partial n_{I,r}}\right)\Big|_{KS-SCF} - \left(\frac{\partial \sigma_I}{\partial n_{I,\sigma}} + \frac{\partial r_I}{\partial n_{I,r}}\right)\Big|_{initial}, \\ &= U_I + \Delta_I \end{aligned} \tag{6}$$

Accordingly, there is a competition (shown as $\Delta_I = (U_{SE} - U_R)_I$ in Eq. (6) above) between two different types of on-site Hubbard-type[30] one-electron potentials (*note*: despite having the same type of interactions, these potentials are not the same as the Hubbard parameter U) obtained by the gradient of self-energy and orbital relaxation with respect to an orbital occupation. As shown in Eq. (6), $U_{SE}$ and $U_R$ are the obtained Hubbard-type potential contribution considered in the LDA+U calculation, which are derived from the linear response calculation of the first-order perturbation of the self-energy and orbital wavefunction relaxation in the orbitals. Therefore, we self-consistent proved our physical idea that *the perturbations of self-energy and wavefunction relaxation with related to the occupation number in the orbital should follow the behavior of Hubbard-type interaction.*

From the perspective of LDA+U within the framework of Cococcioni's linear response method,[24] the corresponding $U_I + \Delta_I$ is thus the self-consistently determined Hubbard-type potential parameter used in the LDA+$U_{SCF}$ calculations to correct the self-energy error and orbital relaxation error, which is originally expressed as a Lagrange multiplier. However, $U_I$ is the degenerated value, as we only consider the degenerated orbital. If $\Delta_I = 0$, the Hubbard-type potential contributions have been well counteracted in the filled shell, as shown in a previous study on self-interaction of the atom with fully occupied valence orbitals by Perdew and Zunger.[31] If $\Delta_I > 0$, there is a residue for partially filled orbitals.

We consider the orbital eigenvalue described by the LDA+U per Anisimov[32] and expressed using the Janak theorem:[29]

$$\varepsilon_i = \partial E / \partial n_i = \varepsilon_{LDA} + U\left(\frac{1}{2} - n_i\right) \quad (7)$$

Here, $n_i$ is the orbital occupation. For fully filled orbitals $(n_i = 1)$, the orbital eigenvalue is shifted by $(-\frac{U}{2})$, while for empty shells $(n_i = 0)$, this value is shifted by $(\frac{U}{2})$. For lanthanides with partially filled f-orbitals $(0 < n_i < 1)$, we cannot simply use the empirical $n_i$ from the valence electron configuration to determine which orbital should be shifted downward or upward. The value of $n_i$ should be calculated *ab initio* and should satisfy the condition of Eq. (6). For 4f fine levels in particular, some of the occupied 4f states need to be shifted downward below the valence band, while some of 4f empty states need to be shifted upward into the forbidden gap or the conduction band on a case-by-case basis. Therefore, Eq. (6) goes the empirical valence electron configuration to give the conditions of the assignment of $n_i$ in the orbitals.

Figure 1 (a)-(c) show examples to clarify the condition set by Eq. (6). For $La^{3+}$ in $La_2O_3$ (Figure 1 (a)), the non-crossover feature reveals a positive residue $\Delta_I > 0$, which means it is actually a partially filled shell with $4f^{\delta}$ (0<δ<1). The $Lu^{3+}$ in $Lu_2O_3$ (Figure 1 (b)) exhibits a perfect crossover feature, meaning that $\Delta_I = 0$ for the $4f^{14}$ orbital. Interestingly, for the example of $Ho^{3+}$ ($4f^{10}$, $n_i$=10/14) in $Ho_2O_3$, the crossover of two lines described by two Hubbard-type contributions of self-energy and orbital relaxation indicates a tendency of fully filled orbital effects through the conditions of Eq. (6), which implies a greater downward shift than dictated by Hund's Rule. This phenomenon will be discussed in a future study.

In crystalline solids, how can we recover the on-site multi-components of orbitals instead of the single-components from the simplified model above? We simply consider the crystal-field splitting (CF) effect, the main effect influencing the Hamiltonian of atoms in crystalline solids. As we know, d orbitals split into two sets, $t_{2g}$ and $e_g$ orbitals, under the influence of CF. Similarly, f orbitals split into three sets: $a_{2u}(f_{xyz})$, $t_{2u}(f_x(y^2-z^2)$, $f_y(z^2-x^2)$, $f_z(x^2-y^2))$, and $t_{1u}(f_{x3}$, $f_{y3}$, $f_{z3})$. Regardless of the detailed assignment of electrons among such orbital components, we simply consider the possibilities of the orbital charge occupation perturbation. For f-orbital based lanthanides, the partially filled f-electrons have three sites that may be perturbed due to the CF effect. Therefore, the perturbed number is $\Delta n = 3\Delta n_0$. Meanwhile, for the fully filled shell, there is only one perturbed possibility that is resonant for all three components: $\Delta n = (1/3)\Delta n_0$.

### (2) Actualization in linear response

Therefore, the second direction of the linear response has been proposed to annihilate the self-induced on-site Hubbard U dipole effects with an elaborately devised Lagrange multiplier for the orbital potential shift for perturbation without any change to the framework of Cococcioni linear response[24] and Janak theorem;[29] this is advantageous for implementation in plane-wave basis set based pseudopotential DFT calculations. This treatment in the present work is *l*-degenerate, where *l* is the maximum angular momentum of the targeted electron orbital. When the effects of these two perturbations are equivalent, the input on-site Coulomb potential $U_{eff}$ is the equivalent self-energy of the electron located in its orbital.

We first refer to the work of Kulik et al.,[33] who used an improved method to eliminate the zero-point error of the linear response of the original procedure of Cococcioni et al.[24] Such perturbation on a given orbital should be independent of the on-site Hubbard U energy. However, the correlated error cannot be strongly reduced in the current method of Kulik et al.[33] This is due to the linear behavior of the output U ($U_{out}$) and input ($U_{in}$), which results in another complicated functional relationship among the variables of the linear response method in terms of orbital perturbation, charge occupancy, and $U_{in}$ because $\partial U_{out}/\partial U_{in}=\partial^2\alpha_I/\partial n\partial U_{in}$=constant. Here, $\alpha_I$ and n are the original orbital perturbation and occupancy, respectively. We proposed splitting the original perturbation into two parts: (1) the perturbation per on-site Coulomb energy and (2) the perturbation of different occupancy levels by the same orbital at the zero on-site Coulomb energy. The two different Lagrange potential shifts applied are given in Eqn. 2, where the details are described in a previous work:[34]

$$\text{set} \quad \begin{cases} \alpha_a = a\cdot\alpha_I = \dfrac{\alpha_I}{U_{in}}\cdot a_0 \\ \alpha_b = U_{in} + \alpha_I \end{cases} \tag{7}$$

From the above, the unit potential is given as $a_0 = 1eV$. We then have two different Lagrange multipliers, one denoting the local-ranged electronic charge occupancy perturbation and another denoting the non-local-ranged magnetic moment perturbation. The Lagrange multipliers are related by

$$\alpha_a\cdot\alpha_b = a_0\cdot\alpha_I + \frac{\alpha_I^2}{U_{in}} \cong a_0\cdot\alpha_I \tag{8}$$

Using the equations proposed by Kulik et al.[33] and the above (Eq. (7)) perturbation, we can reformulate Cococcioni's linear response as

$$\begin{aligned} U_{out1} &= -\left(\frac{\partial\alpha_I}{\partial q(a)} - \frac{\partial\alpha_I^{KS}}{\partial q(a)_{KS}}\right) = -\left(\frac{\partial\alpha_a}{\partial q(a)} - \frac{\partial\alpha_a^{KS}}{\partial q(a)_{KS}}\right)\cdot\left(\frac{U_{in}}{a_0}\right) \\ &= \left(\frac{U_{in}}{a_0}\right)\left(U_{scf1}(U_{in}) - \frac{U_{in}}{m}\right) \end{aligned} \tag{9}$$

$$\begin{aligned} U_{out2} &= -\left(\frac{\partial\alpha_I}{\partial q(b)} - \frac{\partial\alpha_I^{KS}}{\partial q(b)_{KS}}\right) = -\left(\frac{\partial\alpha_b}{\partial q(a)} - \frac{\partial\alpha_b^{KS}}{\partial q(a)_{KS}}\right) \\ &= U_{scf2}(U_{in}) = \left(\frac{a_0}{U_{in}}\right)\cdot U_{scf1}(U_{in}) = \left(\frac{a_0}{U_{in}}\right)\cdot U_{scf}(U_{in}) \end{aligned} \tag{10}$$

Borrowing the concept of variation theory, we write

$$\left|U_{out2} - U_{out1}\right|_{U_{in}} = \left|U_{scf}\left(\frac{a_0}{U_{in}} - \frac{U_{in}}{a_0}\right) + \left(\frac{U_{in}}{a_0}\right)\left(\frac{U_{in}}{m}\right)\right|_{U_{in}} \geq 0 \tag{11}$$

$U_{out1}$ and $U_{out2}$ are found to be equivalent if the minimum value of zero is obtained for Eq. (10). We then have

$$U_{scf}\left(\frac{a_0}{U_{in}}-\frac{U_{in}}{a_0}\right)+\left(\frac{U_{in}}{a_0}\right)\left(\frac{U_{in}}{m}\right)=0 \tag{12}$$

In this way, we obtain $U_{scf}$ in relation to $U_{in}$ at the crossover point between the $U_{out1}(U_{in})$ and $U_{out2}(U_{in})$ plots.

$$U_{scf}=\frac{\left(\frac{U_{in}}{m}\right)}{\left(1-\left(\frac{a_0}{U_{in}}\right)^2\right)} \tag{13}$$

If Eq. (12) does not reach zero, such as in the case of a partially filled shell, there may exist a rigid difference between the two different types of perturbation:

$$\left|U_{out2}-U_{out1}\right|_{U_{in}}=\left|U_{scf}\left(\frac{a_0}{U_{in}}-\frac{U_{in}}{a_0}\right)+\left(\frac{U_{in}}{a_0}\right)\left(\frac{U_{in}}{m}\right)\right|_{U_{in}}=\Delta>0 \tag{14}$$

In this case, the first variation of the equation Eq. (14)=0 and its second variation need to be positive, according to

$$\delta\left(U_{out2}-U_{out1}\right)_{U_{in}}=\delta\left[U_{scf}\left(\frac{a_0}{U_{in}}-\frac{U_{in}}{a_0}\right)+\left(\frac{U_{in}}{a_0}\right)\left(\frac{U_{in}}{m}\right)\right]_{U_{in}}=0 \tag{15}$$

We therefore further obtain

$$\begin{aligned}&\frac{\partial}{\partial U_{in}}\left[U_{scf}\left(\frac{a_0}{U_{in}}-\frac{U_{in}}{a_0}\right)+\left(\frac{U_{in}}{a_0}\right)\left(\frac{U_{in}}{m}\right)\right]_{U_{in}}=0\\&\frac{\partial^2}{\partial U_{in}^2}\left[U_{scf}\left(\frac{a_0}{U_{in}}-\frac{U_{in}}{a_0}\right)+\left(\frac{U_{in}}{a_0}\right)\left(\frac{U_{in}}{m}\right)\right]_{U_{in}}>0\end{aligned} \tag{16}$$

If we choose the constant of above Eq. (16) for a non-fully filled shell, we then obtain a simplified representation of $U_{scf}$ given by:

$$U_{scf}=\frac{\left(\frac{2U_{in}}{m}\right)}{\left(1+\left(\frac{a_0}{U_{in}}\right)^2\right)}+\Delta \tag{17}$$

Then, considering two different effective degeneracies ($m_1$ and $m_2$) for fully filled and partially filled shells, we summarize the two cases above for a self-consistent determination and set the input $U_{in}$ to determine $U_{scf}$ in the DFT+U calculations:

$$U_{scf} = \begin{cases} \dfrac{\left(\dfrac{U_{in}}{m_1}\right)}{\left(1-\left(\dfrac{a_0}{U_{in}}\right)^2\right)} & (occ = 1, crossover) \quad \left(m_1 = \dfrac{1}{l}\right) \\ \\ \dfrac{\left(\dfrac{2U_{in}}{m_2}\right)}{\left(1+\left(\dfrac{a_0}{U_{in}}\right)^2\right)} + \Delta & (non-crossover) \quad \left(m_2 = \left[(2l+1) - \dfrac{occ}{2}\right]\right) \end{cases} \tag{18}$$

We straightforwardly determine *$m_1$* and *$m_2$* in a simple form, such as that described above. The $occ$ is the self-consistent determined electron occupancy in the localized orbital at the zero point energy before the perturbation. According to Eq. (18), the effective degeneracy is changed for the cases of crossover and non-crossover. We therefore divide the physical or chemical problems of fully filled and partially filled orbitals into crossover and non-crossover scenarios. These scenarios do not necessarily correspond to each other, making the linear-response perturbation calculations more convenient. The crossover can occur for a fully filled orbital, such as the $4f^{14}$ orbital of Lu, or a partially filled orbital, such as $2p^4$ for O or even $2p^3$ for N. We illustrate this point using two examples shown in Figures 1(a) and 1(b). We also demonstrate the influence of this new linear response method for the fully filled and partially filled orbitals, as shown in Figure 1(c).

Lany and Zunger have studied the problem of localized oxygen hole states[35-37] in wide band gap semiconductor oxides and proposed treating the problem using a method based on an approach similar to Koopmans' theorem.[21,22,28] We follow their approach and examine the basic determination of the relationship between the system electron energy and its integer occupation number. As shown in Figure 1(d), a correct description of the localized states depends crucially on $d^2E/dn^2$.[38,39] This second derivative of E with respect to the continuous occupation number is concave ($d^2E/dn^2 < 0$) for HF theory but convex for LDA/GGA ($d^2E/dn^2 > 0$). However, the behavior is actually linear,[38-40] such that $d^2E/dn^2 = 0$. Therefore, our proposed method can be described as

$$\left|U_{out2} - U_{out1}\right|_{U_{in}} = \frac{d^2E}{dn^2} = 0 \tag{19}$$

The E in Eq. (19) is not the system total energy. Rather, it is the electronic energy for integer occupation numbers.

**Calculation setup**

The proposed method can be efficiently implemented if the structure is treated as a periodic unit cell for use with the plane-wave basis set. We modeled all 15 lanthanide sesquioxides using the hexagonal (A-type) lattice with the $P\overline{3}m1$ space group. We used the CASTP code to perform our DFT+U calculations.[41] The Ln and O norm-conserving pseudopotentials are generated using the OPIUM code in the Kleinman-Bylander projector form,[42] and the non-linear partial core correction[43] and a scalar relativistic averaging scheme[44] are used to treat the spin-orbital coupling effect. We treated the (4*f*, 5*s*, 5*p*, 5*d*, 6*s*) states as the valence states of the Ln atoms. The RRKJ method is chosen for the optimization of the pseudopotentials.[45]

The PBE functional was chosen for PBE+U calculations with a kinetic cutoff energy of 750 eV, with the valence electron states expressed in a plane-wave basis set. The ensemble DFT (EDFT) method of Marzari et al.[46] is used for convergence. Reciprocal space integration was performed using *k*-point grids of 7 × 7 × 4 *k* points in the $Ln_2O_3$ Brillouin zone. For this *k*-point density, the total energy is converged to less than $5.0\text{x}10^{-7}$ eV per atom. The Hellmann-Feynman forces on the atom were converged to less than 0.01 eV/Å.

We use the Anisimov type DFT+U method[32] and the Hubbard U parameter self-consistently determined for the Ln 4f orbital by our new linear response method. To stabilize the hole states that lie in the O 2p orbitals, we also apply a self-consistently determined Hubbard U potential (method used above) to the O 2p states following the approach of Lany[21,22], Morgan et al.,[47] and Keating et al.[23] Accordingly, both the f- and p-orbital electrons of the rare earth and oxygen atoms should be considered when using DFT+U.[12]

As is well-known, norm-conserving pseudopotentials can reflect all-electron behavior for outer shell valence electrons for **|S-matrix|**=1, unlike the ultrasoft pseudopotentials.[48,49] Therefore, the non-linear core corrected norm-conserving pseudopotential can provide a better response in DFT+U calculations, especially for the calculations of defects.[12] We note that our method actually provides almost identical values of the U parameter for both norm-conserving and ultrasoft pseudopotentials. This means that the obtained value has an intrinsic physical meaning for the studied materials.

## Result and Discussion

Lanthanide sesquioxides have three stable structures in the ambient environment, namely, a hexagonal structure (A-type) with $P\overline{3}m1$ symmetry, monoclinic structure (B-type) with C12/m1 symmetry, and cubic bixbyite structure with $Ia\overline{3}$ symmetry.[20] Light lanthanide sesquioxides usually have A-type lattices, such as $La_2O_3$, $Ce_2O_3$, $Pr_2O_3$, and $Nd_2O_3$, while heavy lanthanide sesquioxides prefer C-type structures. The ground states of all lanthanide sesquioxides exhibit anti-ferromagnetic (AFM) orderings that occur below the Neel temperature (4 K). To simplify our calculations and conclusions for such oxides, we constrain them to A-type lattice structures with AFM ordering to more clearly demonstrate the f-level variation behaviors. The crystal structures used in our modeling are in fact the same as in the work by Gillen et al.[15] and Jiang et al.[14] Prior to our linear response determination of the Hubbard U, the geometries and lattice parameters of all lanthanide sesquioxides were optimized using PBE functional calculations. This procedure reduces the computational cost and ensures the reliability of the Hubbard U value obtained by our self-consistent iterative calculations. We use this procedure prior to the Hubbard U determination because DFT has been already verified to be reliable for the structural optimization of compound solids with 4f or 5f orbitals.[50] This may be due to the well-developed pseudopotential technique[12,50] and, more importantly, to the fact that f-electrons have a small influence on the lattice parameters when treated as valence electrons, as shown by the small difference of the DFT and DFT+U calculated lattice parameters.[12,23,51] Nevertheless, the U parameter must be determined more carefully.[12,23,51]

Figure 2(a) summarizes the calculated band gaps across the $Ln_2O_3$ (Ln=La,…,Lu) series and compares these values to the experimental data of Prokofiev et al. This comparison shows the flexibility of the Hubbard U calculations for DFT+U in band structure calculations, as the hybrid functional and GW methods exhibit some deviation from the experimental values. However, the errors in the values obtained using hybrid functionals are in fact acceptable.[15] Checking our benchmark diagram (Figure 2), we found that with the exception of such oxides as $Ce_2O_3$, $Pr_2O_3$, $Eu_2O_3$, $Dy_2O_3$ and $Lu_2O_3$, the $GW_0$@LDA+U results are quite close to the experimental values.[14] We attribute the errors for these oxides to the inability of the $GW_0$@LDA+U method to perfectly describe the highest occupied and lowest unoccupied f-levels. This is supported by the presence of states that deeper in the forbidden gap, which is the signature of the $f^{1\text{-}2}$ (early light lanthanides), half-filled and fully filled lanthanide solids. To improve these results, more computational effort is required to perform self-consistent GW calculations instead of combining GW with LDA+U, as the empirical LDA+U simply cannot give the correct f-levels, which are then hardly improved by the GW calculations.

We next consider the results obtained using hybrid functionals, such as sX-LDA,[15] shown in Figure 2 (a), with the electronic band gaps calculated for a series of materials from $La_2O_3$ to $Gd_2O_3$ and for $Yb_2O_3$ and $Lu_2O_3$. Comparison of the GW and experimental data shows that sX-LDA with 100% of the HF matrix and a 0.76 $bohr^{-1}$ screening length (Thomas-Fermi) can give satisfactory results for the light lanthanide sesquioxides. However, for materials with heavier elements, such as $Gd_2O_3$, larger errors are obtained. For $Lu_2O_3$, the error even reaches 13%. We note that the sX-LDA results will be improved because the calculations described above were conducted on the basis of the simplest Thomas-Fermi screening model, in which the electrons in the solids are treated as a homogenous electron gas. The constant, 0.76 $bohr^{-1}$ screening length is actually the value for conventional s-p valence electron based semiconductors. However, the enhance-function in the sX-LDA is actually based on Eq. (4) of the Kleimann-Bylander type empirical function in Ref. 39.[52] Therefore, the DFT+U method can be used for precursor calculations to provide a theoretical data reference for further evaluation by advanced sX-LDA methods, as mentioned in previous work.[12] On the other hand, a lower percentage of HF with a smaller screening length, such as that used in the HSE06 function in the work of Gillen et al.,[15] actually provides better results than sX-LDA. This finding suggests that efforts should be made to combine the parameters of the HF matrix and Thomas-Fermi screening length obtained from the "mass-center" of the contour in Figure 2(b) to improve the electronic structure calculation results for lanthanide sesquioxides.

The DTF+U predicted unit cell volumes for the $Ln_2O_3$ series shown in Figure 3 provide a clear confirmation of the "lanthanide contraction". The reduction of the ionic radius of the solids induced by the 4f screening effect is apparent in the contraction of the total unit cell volume from lanthanum (La) to lutetium (Lu). It is interesting that there are actually three jumps in the data in Figure 3 for La, Gd, and Lu, corresponding to empty, half-filled, and fully filled occupancies. This orbital energy trend is similar to that given by the empirical predictions of Nugent,[53] which provide the ground-state interelectronic-repulsion-stabilization energy for all $Ln^{3+}$ (Ln=La,…, Lu) configurations. For the optimized A-type lattice constants, we obtained errors of less than 1% for $La_2O_3$ and $Ce_2O_3$ compared to the experimental data.[54,55] Therefore, our method is a reliable tool for correctly determining the lanthanide crystal structures under extreme physical environments.

We further show that the Hubbard U of the 4f orbital actually exhibits a trend similar to the experimental band gaps. However, the energy differences are too large. Therefore, we use numerical fitting to relate the Hubbard U and experimental band gaps, as shown in Figure 4. The numerical fitting formula is given by:

$$E(U) = \frac{(2l+1)}{2} + \frac{l^2}{4}\ln[\ln(U)] \tag{15}$$

The value of $l$ in the Eq. (15) is actually $l = 3$. This value may be a good reference for future semi-empirical fitting procedures for other theoretical modeling studies.

The observed electronic band gap variation exhibits very similar trends for both the first and second half of the series from La to Lu. Compared to the electronic band gaps, the Hubbard U values show more variation (Figure 4). There is an obvious periodic feature for the U of 4f orbitals, with periodic dips at $Ce_2O_3$, $Tb_2O_3$, and $Yb_2O_3$, except for $La_2O_3$ and $Lu_2O_3$. The $Sm_2O_3$-$Eu_2O_3$-$Gd_2O_3$ and $Tm_2O_3$-$Yb_2O_3$-$Lu_2O_3$ transitions exhibit nearly identical behavior to the Hubbard U variations, and the electronic band gaps also display similar trends. The $Ce_2O_3$-$Sm_2O_3$ and $Tb_2O_3$-$Yb_2O_3$ transitions are slightly different for the Hubbard U trends, as shown in Figure 4, with the light lanthanide sesquioxides having larger values on average than the heavy sesquioxides from $Tb_2O_3$ to $Yb_2O_3$. This latter finding can be interpreted based on the valence electron configurations. Due to the increased 4f electron occupancies among the heavy lanthanide sesquioxides, the 4f electrons actually penetrate deeper into the core shells due to the "lanthanide contraction", and the 4f electron perturbations of the occupancies cannot be higher because of the larger attractive Coulomb interactions between the core and valence 4f electrons. This can be more accurately described using relativistic effects, but this approach is too complicated for our method.[56]

Figure 5(a) shows that the electronic structures of the 4f-orbital based $Ln_2O_3$ are actually very complicated. There are many fine-structure energy levels for the 4f electron transitions. Therefore, it is reasonable that such fine-structure energy levels may play an important role in contributing to the up-conversion luminescence in $Ln^{3+}$ (Ln=La,…, Lu) related materials. We note that Figure 5(a) reveals the fine-structure levels of $Ln^{3+}$ (Ln=La,…, Lu) in the ground states in the A-type sesquioxide crystal lattice. Comparison to the work of Carnall et al.[57] shows that our methods obtains very similar magnitude variation trends.

However, for the electronic band gaps obtained from the band structure calculations, we only focus on the band edges contributed by 4f orbitals, as shown in Figure 5(b). We found that some occupied 4f levels enter into the forbidden band gap arising from the 2p-5d transitions in $Ce_2O_3$, $Pr_2O_3$, $Tb_2O_3$, and $Dy_2O_3$. In other cases, some of the unoccupied 4f levels fall into the 2p-5d transition gap, for example, in $Nd_2O_3$, $Sm_2O_3$, $Eu_2O_3$, and $Yb_2O_3$. The 4f unoccupied levels for $Pr_2O_3$, $Pm_2O_3$ (radioactive), and $Ho_2O_3$ are very close to the 5d band edges and are no more than 0.2 eV higher; in particular, the $Pm_2O_3$ band edge lies just above the 5d edge. Another interesting finding is that $Ce_2O_3$, $Pr_2O_3$, and $Tb_2O_3$ exhibit similar electronic structures. The experimental measurements of Prokofiev et al. indicate four dips in the band gap. Based on the data in Figure 5(b), we now confirm the presence of these dips for $Ce_2O_3$, $Eu_2O_3$, $Tb_2O_3$, and $Yb_2O_3$. We further verified that these gaps occur for $Ce_2O_3$ and $Tb_2O_3$ because the 4f occupied levels of these materials enter the 2p-5d gap; meanwhile, the 4f unoccupied levels for $Eu_2O_3$ and $Yb_2O_3$ fall into the 2p-5d gaps. Taken together, these effects lead to a lower optical transition band gap in optical absorption measurements, similar to the findings of Prokofiev et al.[6]

Finally, we discuss the band structures obtained by our method for the A-type $Ln_2O_3$ (Ln=La,…, Lu), as shown in Figure 6. We found that $Lu_2O_3$ has a direct band gap of 5.5 eV. This is promising for the use of these materials in luminescence applications due to the vertical optical excitation path. However, there is a lack of both theoretical and experimental data relating the optical properties to the defect states of these materials. The gap between the experimental and theoretical studies remains an open question that will be the subject of future studies.

## Conclusion

We have shown that our proposed two-way linear response method can accurately determine the on-site screened Coulomb potential for both cations and anions in the 4f orbital based rare earth metal oxides ($Ln_2O_3$, Ln=La,…, Lu). This method can also determine whether the orbital shell tends to be fully filled and whether it can approach 0 eV. Using this method, we applied our self-consistently determined on-site Coulomb potential for the determination of the Hubbard U parameters of the 4f, 5d and 2p orbitals for the cations and anions. We obtained results in close agreement with the experimental data for the electronic and atomic properties of lanthanide sesquioxides. This work will accelerate the pace of research and discovery in the electronic engineering of next-generation materials that form or are synthesized under extreme physical (new solid phase for materials under high-pressure synthesis process) or chemical environments.

## Acknowledgement

The author, BH, is very grateful to Prof. Chun-Hua Yan and Dr. Hong Jiang of Peking University for many helpful discussions. The author gratefully acknowledges the support of Natural Science Foundation of China (NSFC) for Young Scientists (Grant No. NSFC 11504309).

## References

1. K. Van den Eeckhout, D. Poelman, P. Smet, *Materials* **2013**, *6*, 2789.
2. J. Trojan-Piegza, E. Zych, J. Hölsä, J. Niittykoski, *J. Phys. Chem. C* **2009**, *113*, 20493-20498.
3. M. P. Hehlen, M. G. Brik, K. W. Krämer, *J. Lumin.* **2013**, *136*, 221-239.
4. S. Y. Savrasov, G. Kotliar, E. Abrahams, *Nature* **2001**, *410*, 793-795.
5. P. S. Riseborough, *Adv. Phys.* **2000**, *49*, 257-320.
6. A. V. Prokofiev, A. I. Shelykh, B. T. Melekh, *J. Alloy. Compd.* **1996**, *242*, 41-44.
7. N. V. Skorodumova, R. Ahuja, S. I. Simak, I. A. Abrikosov, B. Johansson, B. I. Lundqvist, *Phys. Rev. B* **2001**, *64*, 115108.

8. D. A. Andersson, S. I. Simak, B. Johansson, I. A. Abrikosov, N. V. Skorodumova, *Phys. Rev. B* **2007**, *75*, 035109.

9. E. Rogers, P. Dorenbos, E. v. d. Kolk, *New J. Phys.* **2011**, *13*, 093038.

10. L. Petit, A. Svane, Z. Szotek, W. M. Temmerman, *Phys. Rev. B* **2005**, *72*, 205118.

11. L. Petit, R. Tyer, Z. Szotek, W. M. Temmerman, A. Svane, *New J. Phys.* **2010**, *12*, 113041.

12. B. Huang, R. Gillen, J. Robertson, *J. Phys. Chem. C* **2014**, *118*, 24248-24256.

13. H. Jiang, R. I. Gomez-Abal, P. Rinke, M. Scheffler, *Phys. Rev. Lett.* **2009**, *102*, 126403.

14. H. Jiang, P. Rinke, M. Scheffler, *Phys. Rev. B* **2012**, *86*, 125115.

15. R. Gillen, S. J. Clark, J. Robertson, *Phys. Rev. B* **2013**, *87*, 125116.

16. N. Sarmadian, R. Saniz, B. Partoens, D. Lamoen, K. Volety, G. Huyberechts, J. Paul, *Phys. Chem. Chem. Phys.* **2014**, *16*, 17724-17733.

17. G. Seguini, E. Bonera, S. Spiga, G. Scarel, M. Fanciulli, *Appl. Phys. Lett.* **2004**, *85*, 5316-5318.

18. S. J. Clark, J. Robertson, *Phys. Rev. B* **2010**, *82*, 085208.

19. J. Robertson, *Eur. Phys. J. Appl. Phys.* **2004**, *28*, 265-291.

20. G.-y. Adachi, N. Imanaka, *Chem. Rev.* **1998**, *98*, 1479-1514.

21. S. Lany, A. Zunger, *Phys. Rev. B* **2009**, *80*, 085202.

22. S. Lany, A. Zunger, *Phys. Rev. B* **2010**, *81*, 205209.

23. P. R. L. Keating, D. O. Scanlon, B. J. Morgan, N. M. Galea, G. W. Watson, *J. Phys. Chem. C* **2011**, *116*, 2443-2452.

24. M. Cococcioni, S. de Gironcoli, *Phys. Rev. B* **2005**, *71*, 035105.

25. L. Bendavid, E. Carter, in First Principles Approaches to Spectroscopic Properties of Complex Materials; C. Di Valentin, S. Botti, M. Cococcioni, Eds.; Springer: Berlin Heidelberg, **2014**; pp 47-98.

26. J. Andriessen, P. Dorenbos, C. W. E. van Eijk, *Phys. Rev. B* **2005**, *72*, 045129.

27. J. Andriessen, P. Dorenbos, C. W. E. van Eijk, *Physical Review B* **2005**, *72*, 045129.

28. A. Zunger, A. J. Freeman, *Phys. Rev. B* **1977**, *16*, 2901-2926.

29. J. F. Janak, *Phys. Rev. B* **1978**, *18*, 7165-7168.

30. J. Hubbard, *Proceedings of the Royal Society of London A: Mathematical, Physical and Engineering Sciences* **1963**, *276*, 238-257.

31. J. P. Perdew, A. Zunger, *Phys. Rev. B* **1981**, *23*, 5048-5079.

32. I. A. Vladimir, F. Aryasetiawan, A. I. Lichtenstein, *J. Phys. Condens. Matter* **1997**, *9*, 767.

33. H. J. Kulik, M. Cococcioni, D. A. Scherlis, N. Marzari, *Phys. Rev. Lett.* **2006**, *97*, 103001.

34. B. Huang, *eprint arXiv:1507.05040* **2015**.

35. G. Pacchioni, F. Frigoli, D. Ricci, J. A. Weil, *Phys. Rev. B* **2000**, *63*, 054102.

36. M. d’Avezac, M. Calandra, F. Mauri, *Phys. Rev. B* **2005**, *71*, 205210.

37. A. Droghetti, C. D. Pemmaraju, S. Sanvito, *Phys. Rev. B* **2008**, *78*, 140404.

38. J. P. Perdew, A. Ruzsinszky, G. I. Csonka, O. A. Vydrov, G. E. Scuseria, V. N. Staroverov, J. Tao, *Phys. Rev. A* **2007**, *76*, 040501.

39. P. Mori-Sánchez, A. J. Cohen, W. Yang, *Phys. Rev. Lett.* **2008**, *100*, 146401.

40. J. P. Perdew, R. G. Parr, M. Levy, J. L. Balduz, *Phys. Rev. Lett.* **1982**, *49*, 1691-1694.

41. S. J. Clark, M. D. Segall, C. J. Pickard, P. J. Hasnip, M. I. J. Probert, K. Refson, M. C. Payne, *Zeitschrift Fur Kristallographie* **2005**, *220*, 567.

42. L. Kleinman, D. M. Bylander, *Phys. Rev. Lett.* **1982**, *48*, 1425-1428.

43. S. G. Louie, S. Froyen, M. L. Cohen, *Phys. Rev. B* **1982**, *26*, 1738-1742.

44. I. Grinberg, N. J. Ramer, A. M. Rappe, *Phys. Rev. B* **2000**, *62*, 2311-2314.

45. A. M. Rappe, K. M. Rabe, E. Kaxiras, J. D. Joannopoulos, *Phys. Rev. B* **1990**, *41*, 1227-1230.

46. N. Marzari, D. Vanderbilt, M. C. Payne, *Phys. Rev. Lett.* **1997**, *79*, 1337-1340.

47. B. J. Morgan, G. W. Watson, *J. Phys. Chem. C* **2010**, *114*, 2321-2328.

48. P. J. Hasnip, C. J. Pickard, *Comput. Phys. Commun.* **2006**, *174*, 24-29.

49. K. Laasonen, A. Pasquarello, R. Car, C. Lee, D. Vanderbilt, *Phys. Rev. B* **1993**, *47*, 10142-10153.

50. C. J. Pickard, B. Winkler, R. K. Chen, M. C. Payne, M. H. Lee, J. S. Lin, J. A. White, V. Milman, D. Vanderbilt, *Phys. Rev. Lett.* **2000**, *85*, 5122-5125.

51. T. Zacherle, A. Schriever, R. A. De Souza, M. Martin, *Phys. Rev. B* **2013**, *87*, 134104.

52. D. M. Bylander, L. Kleinman, *Phys. Rev. B* **1990**, *41*, 7868-7871.

53. L. J. Nugent, *J. Inorg. Nuclear Chem.* **1970**, *32*, 3485-3491.

54. A. Proessdorf, M. Niehle, M. Hanke, F. Grosse, V. Kaganer, O. Bierwagen, A. Trampert, *Appl. Phys. Lett.* **2014**, *105*, 021601.

55. H. Pinto, M. H. Mintz, M. Melamud, H. Shaked, *Phys. Lett. A* **1982**, *88*, 81-83.

56. P. Pyykko, *Chem. Rev.* **1988**, *88*, 563-594.

57. W. T. Carnall, G. L. Goodman, K. Rajnak, R. S. Rana, *J. Chem. Phys.* **1989**, *90*, 3443-3457.

**Table 1.** Comparison of the minimum electronic band gaps (eV) of the lanthanide sesquioxides obtained by theoretical calculations and experimental measurements.

| $Ln_2O_3$ | LDA+U[a] | $GW_0$[a] | sX-LDA[b] | HSE06[b] | this work | Exp.[c] |
|---|---|---|---|---|---|---|
| $La_2O_3$ | 3.76 | 5.24 | 5.50 | 5.10 | 5.53 | 5.5 |
| $Ce_2O_3$ | 2.24 | 1.29 | 1.75 | 3.38 | 2.51 | 2.4 |
| $Pr_2O_3$ | 3.17 | 2.82 | 3.80 | 3.77 | 3.89 | 3.9 |
| $Nd_2O_3$ | 3.69 | 4.70 | 4.65 | 4.63 | 4.88 | 4.7 |
| $Pm_2O_3$ | 3.35 | 5.41 | 5.60 | 4.80 | 5.19 | |
| $Sm_2O_3$ | 2.15 | 5.22 | 4.80 | 3.40 | 5.11 | 5.0 |
| $Eu_2O_3$ | 1.28 | 3.48 | 4.00 | 2.50 | 4.56 | 4.4 |
| $Gd_2O_3$ | 3.58 | 5.30 | 4.85 | 5.26 | 5.46 | 5.4 |
| $Tb_2O_3$ | 3.34 | 3.74 | | 4.00 | 3.80 | 3.8 |
| $Dy_2O_3$ | 3.47 | 4.24 | | 4.90 | 5.02 | 4.9 |
| $Ho_2O_3$ | 3.05 | 5.12 | | 5.08 | 5.33 | 5.3 |
| $Er_2O_3$ | 2.69 | 5.22 | | 5.30 | 5.50 | 5.3 |
| $Tm_2O_3$ | 1.73 | 5.15 | | 4.80 | 5.40 | 5.4 |
| $Yb_2O_3$ | 1.25 | 4.70 | 4.50 | 3.24 | 4.82 | 4.9 |
| $Lu_2O_3$ | 3.18 | 4.99 | 4.85 | 5.14 | 5.58 | 5.5 |

[a]Reference 13.
[b]Reference 14.
[c]Reference 5.

# Figure 1

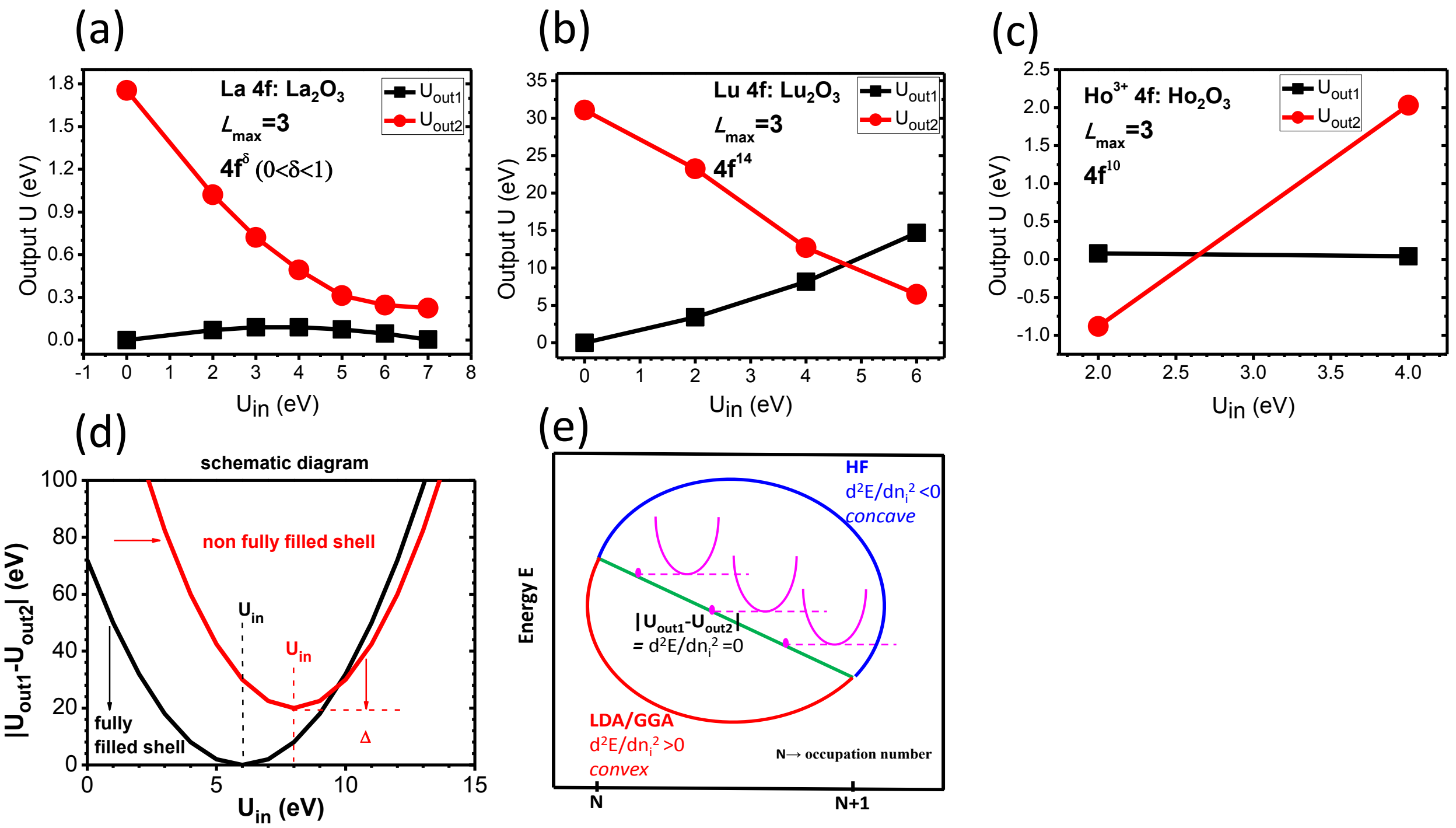

**Figure 1**. (a) Two different outputs of nearly empty 4f shell on-site screened Coulomb potential ($La_2O_3$). (b) Two different outputs of fully filled 4f shell on-site screened Coulomb potential ($Lu_2O_3$). (c) $|U_{out1}$-$U_{out2}|$ vs $U_{in}$ behavior variation for fully filled and non-fully filled shells. (d) Electronic energy vs integer/fractional occupation numbers with different theoretical models for many-body calculations.

**Figure 2**

(a)

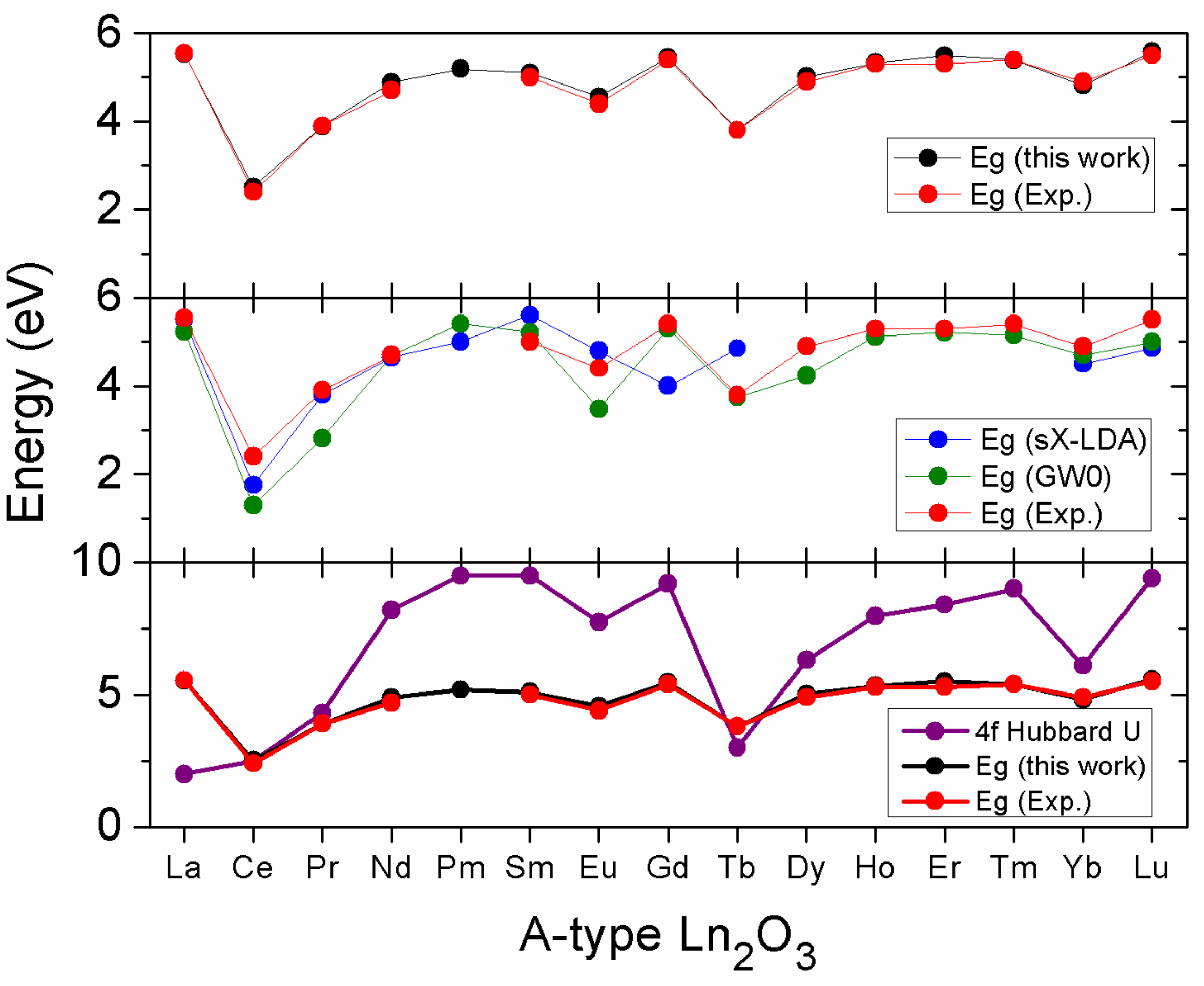

Energy (eV)
Eg (this work)
Eg (Exp.)
Eg (sX-LDA)
Eg (GW0)
Eg (Exp.)
4f Hubbard U
Eg (this work)
Eg (Exp.)
La Ce Pr Nd Pm Sm Eu Gd Tb Dy Ho Er Tm Yb Lu
A-type $Ln_2O_3$

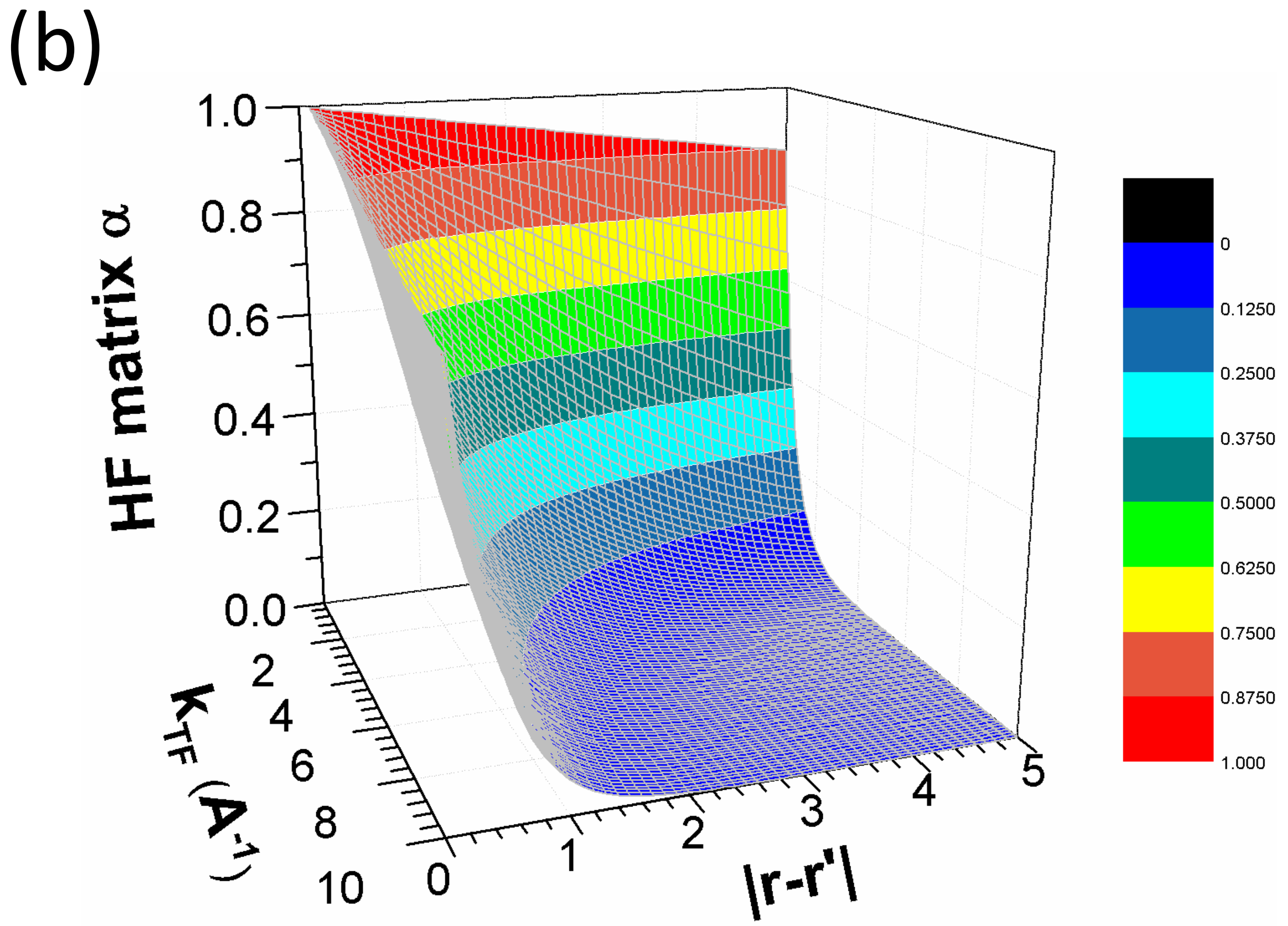

**Figure 2**. (a) Band gaps of $Ln_2O_3$ (from La to Lu) obtained from sX-LDA, $GW_0$ and experimental results as well as the variation trend of the 4f Hubbard U parameters. (b) "Bow-Stern-Contour" variation behavior of both the HF matrix and Thomas-Fermi screening length for sX-LDA methods. (This contour is similar to the bow of a ship).

**Figure 3**

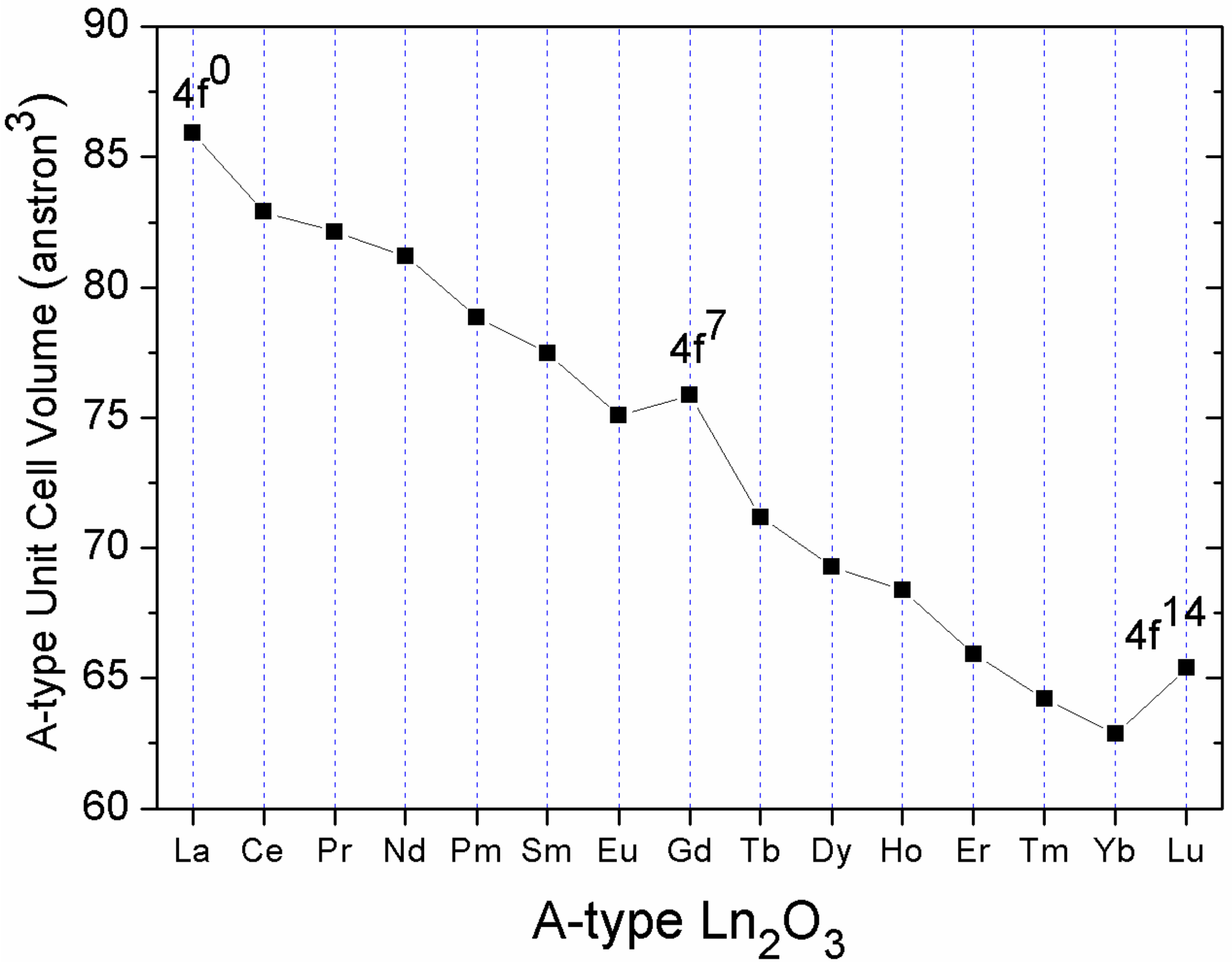

**Figure 3**. Ground-state relaxed A-type unit cell volumes of $Ln_2O_3$ (from La to Lu). The three jumps correspond to empty, half-filled, and fully filled occupancies.

**Figure 4**

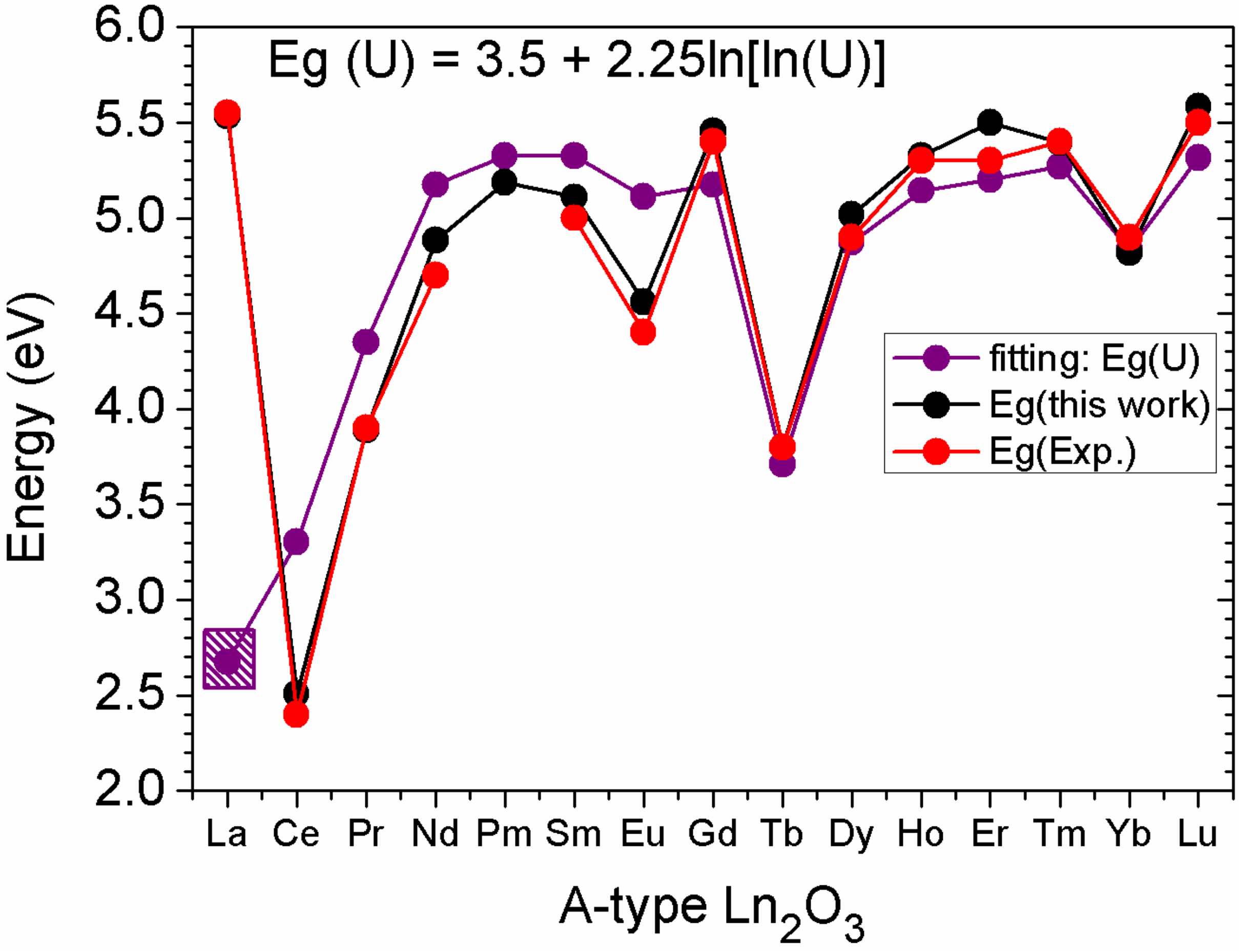

**Figure 4**. Band gap fitting functions (Eq. 30) for different Hubbard U energies and experimental data. The purple shaded area denotes the exclusion of La because its 4f orbital is basically empty, with a small predicted Hubbard U value. The $La_2O_3$ band gap is essentially only due to the O-2p to La-5d optical transitions.

**Figure 5**

(a)

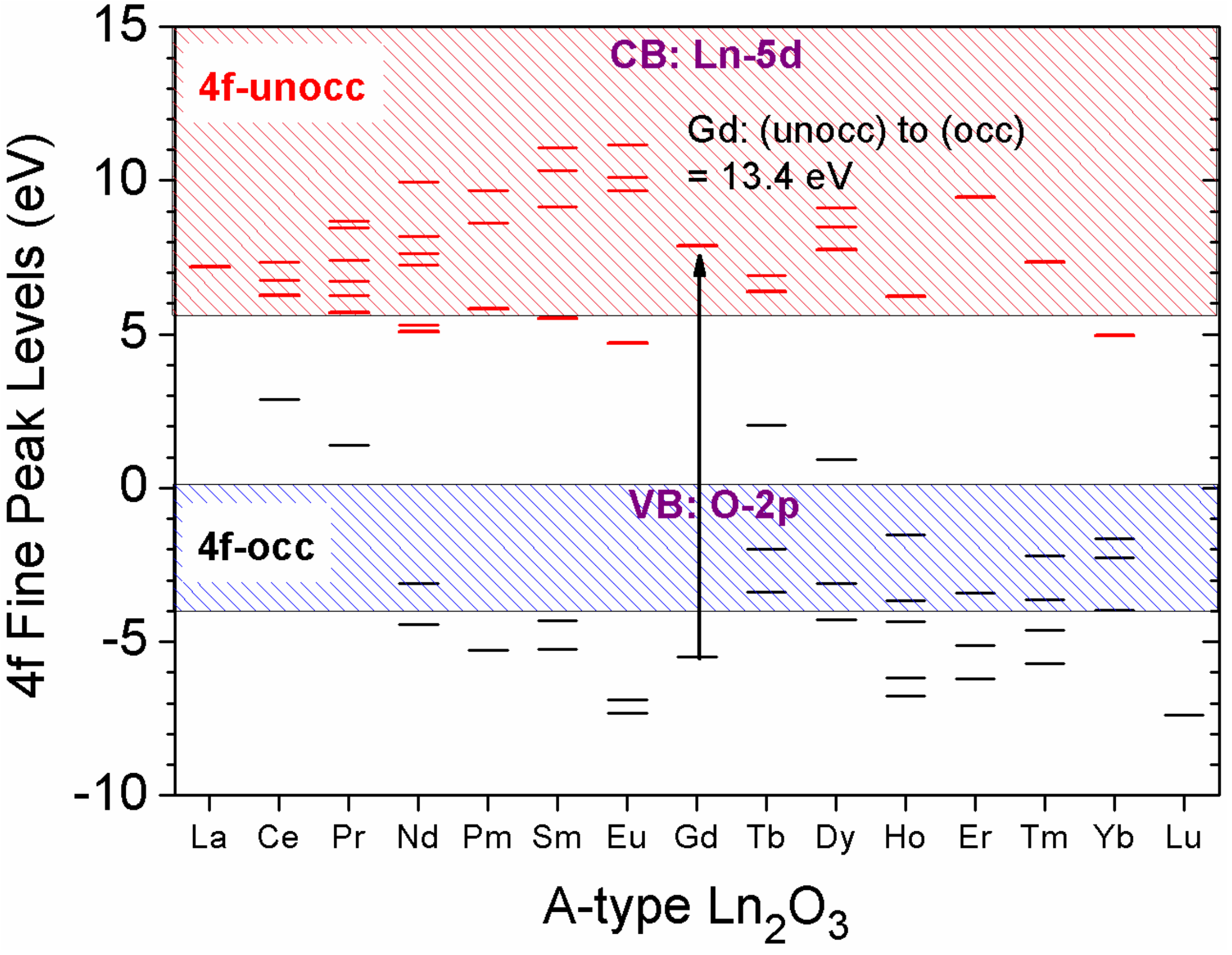
4f-unocc
CB: Ln-5d
Gd: (unocc) to (occ)
= 13.4 eV
VB: O-2p
4f-occ
4f Fine Peak Levels (eV)
15
10
5
0
-5
-10
La Ce Pr Nd Pm Sm Eu Gd Tb Dy Ho Er Tm Yb Lu
A-type $Ln_2O_3$

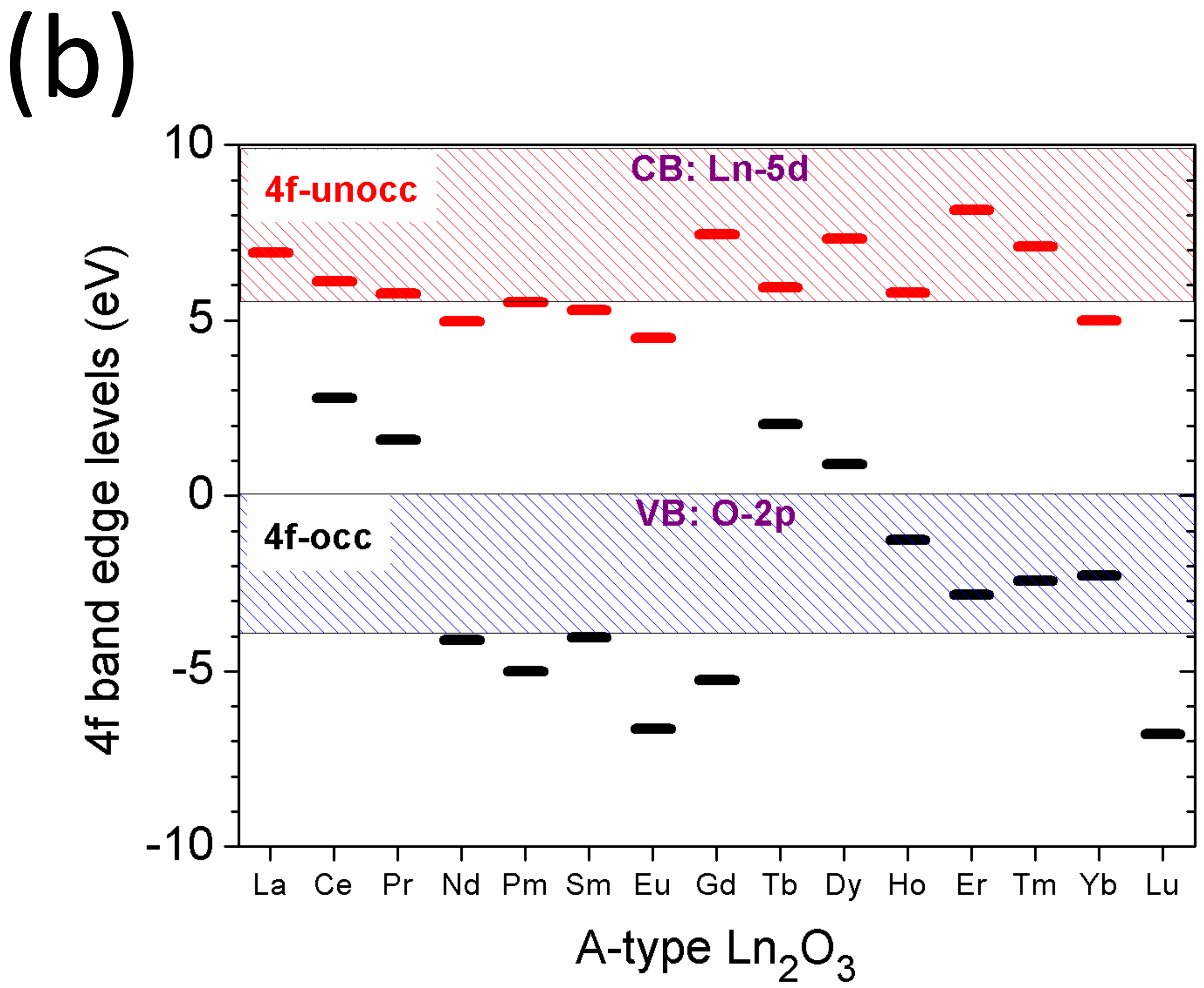

**Figure 5**. (a) Fine-structure energy levels for both occupied and unoccupied 4f orbitals in the ground state. (b) 4f band edge levels for ground-state relaxed A-type $Ln_2O_3$ (from La to Lu).

**Figure 6**

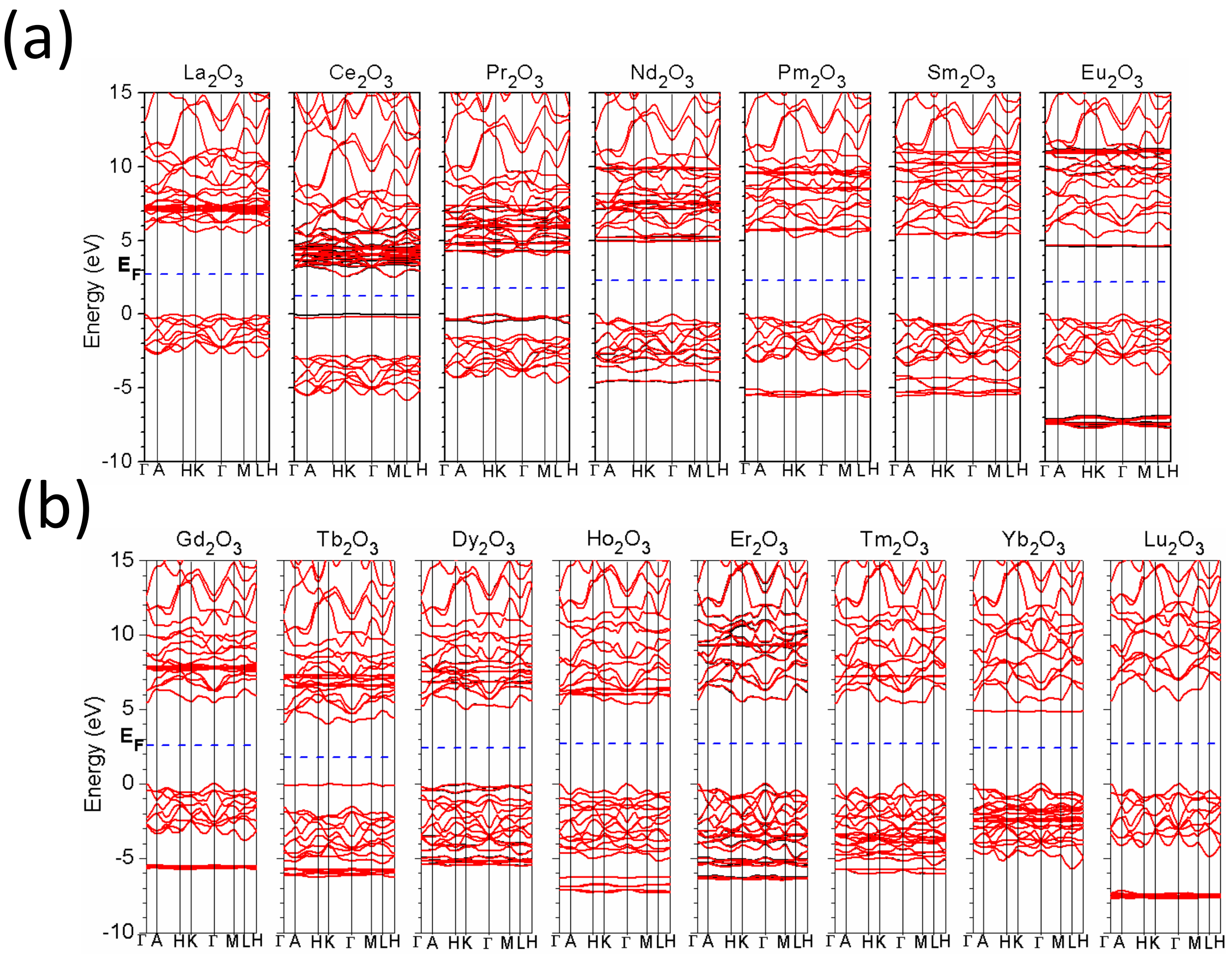

**Figure 6.** (a) Electronic band structures of light $Ln_2O_3$ (from La to Eu). (b) Electronic band structures of heavy $Ln_2O_3$ (from Gd to Lu). Blue dashed lines show the Fermi levels ($E_F$).